\documentclass[a4paper]{jpconf}
\usepackage{graphicx}

\usepackage{amsmath}
\usepackage{graphics}
\usepackage{graphicx}
\usepackage{epsfig}

\usepackage[english]{babel}
\usepackage{color}
\usepackage{verbatim}

\definecolor{azul}{rgb}{0.1,0.2,0.6} 
\definecolor{verde}{rgb}{0.1,0.5,0.3}
\definecolor{bordo}{rgb}{0.9,0.3,0.3}
\usepackage{lipsum}
\usepackage{caption}

\usepackage{amsfonts}
\usepackage{verbatim}
\usepackage{color}
\usepackage{epsfig}
\usepackage{amsfonts}
\usepackage{amssymb}
\usepackage{amsmath}
\usepackage{amscd}
\usepackage{amsthm}
\usepackage{epic}
\usepackage[T1]{fontenc}
\usepackage[utf8]{inputenc}
\usepackage[basic]{circ}
\usepackage{graphics}
\usepackage{cite}
\usepackage[english]{babel}

\newtheorem*{1 Kir}{1$^{\circ}$ Ley de Kirchhoff}
\newtheorem*{2 Kir}{2$^{\circ}$ Ley de Kirchhoff}

\definecolor{azul}{rgb}{0.1,0.2,0.6} 
\definecolor{verde}{rgb}{0.1,0.5,0.3}
\definecolor{bordo}{rgb}{0.9,0.3,0.3}
\usepackage{lipsum}

\usepackage{tikz}
\usetikzlibrary{automata}
\usetikzlibrary{topaths}
\usepackage{pstricks}
\usetikzlibrary{automata,positioning,calc}
\usetikzlibrary{arrows}

\usepackage{framed}

\begin{document}


\title{THE QUANTUM CP-VIOLATING KAON SYSTEM REPRODUCED IN THE ELECTRONIC LABORATORY \\ \vspace {0.5cm} {\small HOMAGE TO NOLBERTO MARTINEZ}\vspace {0.5cm}}

\author{M. Caruso}%
\address{Laboratorio de F\'{\i}sica Te\'orica, Departamento de F\'{\i}sica,  Facultad de Ciencias Exactas, Universidad Nacional de La Plata; IFLP-CONICET, C.C. 67, 1900 La Plata, Argentina.}
\ead{mcaruso@ugr.es}

\author{H. Fanchiotti}%
\address{Laboratorio de F\'{\i}sica Te\'orica, Departamento de F\'{\i}sica,  Facultad de Ciencias Exactas, Universidad Nacional de La Plata; IFLP-CONICET, C.C. 67, 1900 La Plata, Argentina.}

\author{C.A. Garc\'ia Canal}%
\address{Laboratorio de F\'{\i}sica Te\'orica, Departamento de F\'{\i}sica,  Facultad de Ciencias Exactas, Universidad Nacional de La Plata; IFLP-CONICET, C.C. 67, 1900 La Plata, Argentina.}

\author{M. Mayosky}%
\address{LEICI, Departamento de Electrotecnia  Facultad de Ingenier\'ia, Universidad Nacional de La Plata, La Plata, Argentina.  Comisi\'on de Investigaciones Científicas de la Provincia de Buenos Aires-CICpBA, Argentina}

\author{A. Veiga}%
\address{LEICI, Departamento de Electrotecnia  Facultad de Ingenier\'{\i}a,Universidad Nacional de La Plata, La Plata, Argentina; CONICET}

\begin{abstract}
The equivalence between the $\mathrm{Schr\ddot{o}dinger}$ dynamics of a quantum system with a finite number of basis states  and a classical dynamics is realized in terms of electric networks. The isomorphism that connects in a univocal way both dynamical systems was applied to the case of neutral mesons, kaons in particular, and the class of electric networks univocally related to the quantum system was analyzed. Moreover, under $CPT$ invariance, the relevant $\epsilon$ parameter that measures $CP$ violation in the kaon system is reinterpreted in terms of network parameters. All these results were explicitly shown by means of both a numerical simulation of the implied networks and by constructing the corresponding circuits.
\end{abstract}

\section{Introduction}

After the proposal of Ref.\cite{Rosner} (see also \cite{Rosner2,Kostelecky-Roberts}) of an analogy between the physics of the  weak decay of neutral K-mesons (kaons) and a classical system of  oscillators either electrical \cite{Rosner} or mechanical \cite{Rosner2,Kostelecky-Roberts,Reiser:2012fh,Schubert:2011hd}, it was shown \cite{Caruso:2011tu} that this analogy is an equivalence, stricto sensu from  the mathematical point of view. This equivalence is an isomorphism that connects in a univocal way the $\mathrm{Schr\ddot{o}dinger}$ dynamics of a quantum system with a finite number of basis states  and a classical dynamics.

As already stated in \cite{JonaLasinio}, analogies have an important impact in the development of theoretical physics. They may be similarities of physical concepts related to similarities in the mathematical formalization or it may be a purely mathematical equivalence to suggest the development of analogous physical concepts.

This paper presents the construction, via electronic circuits, of the classical equivalent system to a quantum system. In particular,
the well known oscillatory behavior between particle and antiparticle that neutral mesons present is quantitatively reproduced. In this case of neutral kaons, one is interested in the aspects of \textit{Charge conjugation}$-$\textit{Parity}, $CP$, invariance \cite{Lee}. In the context of validity of $CPT$ symmetry, the equivalent analysis of \textit{Time reversal}, $T$, invariance can be considered.

The class of electric networks $\mathcal{R}$ considered is univocally related to the kaon system because one finds the complete map between the matrix elements of the effective Hamiltonian of kaons and those elements of the classical dynamics of the networks. Moreover, there exists a one to one relationship between the states $|K^0\rangle$ and $|\bar{K}^0\rangle$ and port voltages, or currents, of the electric network.

Following this lines we can give a formal classical test of the $CP$ invariance that is a reflection of the corresponding quantum test. One also concludes that any violation of the  $CP$ (or $T$) symmetry is  directly related to the non-reciprocity of the network \cite{Rosner}. In fact, the observable related to the violation of $T$ invariance at quantum level is associated the conductance of a non-reciprocal element needed to be included in the network, the gyrator. This is a two ports, non-reciprocal, passive network without losses that violates the classical symmetry $T$ \cite{Bala,Carlin,Tellegen}. In this way, one ends up with a network completely equivalent to the kaon system, that allows one to present the relevant parameters of the quantum system in terms of circuit parameters. The interaction between both initial subnetworks gives rise to a shift in the proper initial free frequencies, in the same way as the masses of kaons. Moreover, the presence of proper relaxation times of the circuit are associated the mean lives of the combinations $K-$\textit{short} and $K-$\textit{long}. The purpose of this paper is to transcend the formal aspects introduced in \cite{Caruso:2011tu}, presenting details not only of the numerical simulation of the previously proposed circuits but also to explicitly show the implementation of the circuit together with the corresponding experimental measurements.

In Section 2 we briefly summarize the equivalence between the quantum and the classical dynamics, in particular for the case of the neutral kaon system. This section also includes a brief account of the electric networks of interest. Section 3 summarizes the  physical observables in both systems. The design, simulation and realization of the electric circuit is presented in Section 4. Finally, in Section 5 we state our conclusions.

\section{Equivalence Between Dynamics} \label{cap1}

\subsubsection*{Kaons and Oscillators}

Let us consider a quantum system $\mathcal{Q}$ of $\textit{n=2}$ orthonormal basis states denoted by $\{|j\rangle:\, j=1,2\}$ in a certain Hilbert space driven by a Hamiltonian  $\mathbf{H}$.

The system is described by a vector $\pmb{\psi}(t)$ on $\mathbb{C}^2$ that can be written as
$\pmb{\psi}(t)= \textbf{(}\psi_1(t),\psi_2(t)\textbf{)}^\intercal$ in terms of the coordinates $\psi_j(t)$=$\langle j|\psi(t)\rangle$. This  $\pmb{\psi}(t) $ satisfies the  $\mathrm{Schr\ddot{o}dinger}$ equation
\begin{equation}\label{Sch}
\imath d_t|\psi(t)\rangle = \mathbf{H} |\psi(t)\rangle\,\,\,\,or\,\,\,\,  d_t\pmb{\psi}(t)=\mathbf{K}\:\pmb{\psi}(t)
\end{equation}
where $\mathbf{K} \in\mathbb{C}^{2\mathrm{x}2}$, with elements $\mathrm{K}_{ij}=-	\imath\langle i|\mathbf{H}|j\rangle$.

In order to correctly state the equivalence with a classical system it is necessary to perform a decomplexification \cite{Caruso:2011tu}. 
Consequently, the vector decomplexification map $\mathfrak{D}:\mathbb{C}^{2}\longrightarrow\mathbb{R}^{4}$, gives rise to $\mathfrak{D}(\boldsymbol{\psi})= \textbf{(}\pmb{\varphi}_1,\pmb{\varphi}_2\textbf{)}^\intercal$ with $\pmb{\varphi}_1 =  \textbf{(}\Re(\psi_1),\Re(\psi_2)\textbf{)}^\intercal$, $\pmb{\varphi}_2 = \textbf{(}\Im(\psi_1),\Im(\psi_2)\textbf{)}^\intercal$ and $^\intercal$ denotes the matrix transposition. We use the decomplexification introduced by Arnold \cite{Arnold2} that is equivalent to the process presented in \cite{Caruso:2011tu}. Equation \eqref{Sch} can be written as 
\begin{equation}
d_t \left( \begin{matrix}
\pmb{\varphi}_1\\
\pmb{\varphi}_2
\end{matrix}\right)
=
\left( \begin{matrix}
 \mathbf{K}_r   &-\mathbf{K}_i\\
 \mathbf{K}_i    &\hspace*{0.3cm} \mathbf{K}_r
\end{matrix}\right)
\left( \begin{matrix}
\pmb{\varphi}_1\\
\pmb{\varphi}_2
\end{matrix}\right)
\end{equation}
with $\mathbf{K}_r$ and $\mathbf{K}_i$ being the real and the imaginary part of $\mathbf{K}$, respectively. 
The non-hermitian character of $\mathbf{H}$ 
is in order because the kaons decay.
After a standard decoupling procedure one gets the equations
\begin{equation} \label{soe}
\pmb{\ddot{\varphi}}_j(t) - 2\,\mathbf{K}_r \,\pmb{\dot{\varphi}}_j(t) +
(\mathbf{K}_r^2 + \mathbf{K}_i^2)\,\pmb{\varphi}_j(t)=\pmb{0}.
\end{equation}
It is clear that even if both the real and imaginary part of $\pmb{\psi}$ verify the same equation, one cannot leave out one of them because the solution of Eq.\eqref{soe} implies the knowledge of the initial conditions. In this second order case, one  needs to specify the function and the first derivative at $t=0$, while in quantum mechanics one only knows the function. In order to fix the first derivative at $t=0$, $d_t\pmb{\psi}(0)$, one needs the knowledge of $\pmb{\psi}(0)$ and of $\mathbf{K}$ from \eqref{Sch}. Moreover, the calculation of the probability density $|\pmb{\psi}(t)|^2$ necessarily includes both real and imaginary parts.

Let us now go to a classical system $\mathcal{C}$. We start with a
system of linear differential equations of second order, entirely similar to Eq.\eqref{soe}
\begin{equation}\label{A B}
\ddot{\pmb{q}}(t)+\mathbf{A}\:\dot{\pmb{q}}(t)+\mathbf{B}\:\pmb{q}(t)=\pmb{0}
\end{equation}
with $\pmb{q}:\mathbb{R}\longrightarrow\mathbb{R}^{2}$ are the generalized coordinates; $\mathbf{A}, \mathbf{B}\in\mathcal{M}_{2\times 2}(\mathbb{R})$.

The equivalence (isomorphism) between $\mathcal{Q}$ and $\mathcal{C}$ dynamics, discussed in detail in Ref.\cite{Caruso:2011tu}, implies that the real part and the imaginary part of the quantum function are each associated a real classical system. One can eventually take two identical classical systems but prepared with different initial conditions.

We see that is possible establish a \textit{bridge} between this two systems of two states $(|1\rangle,|2\rangle)$ and $(q_1,q_2)$ via the isomorphism $\mathit{\Phi}$ presented in \cite{Caruso:2011tu}. This bridge can be established to translate (as a dictionary) two systems with any number of denumberable states. 
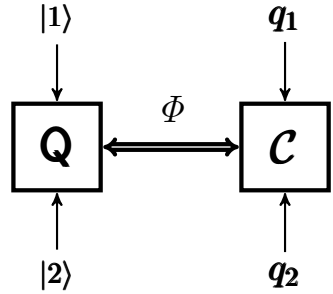
\begin{figure}[!h]
\begin{center}
\begin{tikzpicture}[->,>=stealth', auto, thick, node distance=2.5cm]
\tikzstyle{every state}=[fill=white,draw=black,ultra thick,text=black,scale=1.2]
\node[state,rectangle]  (A) {\large $\pmb{\mathsf{Q}}$};
\node[state,rectangle]  (B)[right of=A]   {\large $\pmb{\mathcal{C}}$};
\node    (1)[xshift=0cm,yshift=1.7cm]  { $\pmb{|1\rangle}$};
\node    (2)[xshift=0cm,yshift=-1.7cm] { $\pmb{|2\rangle}$};
\node    (V1)[xshift=3cm,yshift=1.7cm]  {\large $\pmb{q_1}$};
\node    (V2)[xshift=3cm,yshift=-1.7cm] {\large $\pmb{q_2}$};
\node    (Phi)[xshift=1.5cm,yshift=0.5cm] {\large $\mathit{\Phi}$};
\draw[ultra thick,implies-implies, double equal sign distance] (A) -- (B);
\path
(1)  edge[above] (A)
(2)  edge[above] (A)
(V1) edge[above] (B)
(V2) edge[above] (B);
\end{tikzpicture}
\end{center}
\caption{\footnotesize{Schematic idea to establish a map between this two dynamics. The symbol $\pmb{\mathsf{Q}}$ represent the \textit{deterministic part} of the quantum system $\mathcal{Q}$, i.e. only its Hamiltonian time evolution given by \eqref{Sch}.}}
\label{esquema}
\end{figure}

We will see below that in the case of electrical networks this is related with different voltages in each case, representing the real and the imaginary parts of $\pmb{\psi}$.

We are particularly interested in the quantum system of neutral kaons, because, under the hypothesis of Wigner-Weisskopf \cite{Wigner-Weisskopf}, it can be written as a two-state system. This exemplifies very easily the equivalence with a classical system.

General principles on the basis of quantum field theory guarantee the validity of the $CPT$ symmetry \cite {Luders}. Consequently, in this context is equivalent to speak about $CP$ or $T$ invariance, or non-invariance. Notice that when $CPT$ is a symmetry, the masses of a particle and its antiparticle have to be equal \cite{Lee}.

We consider here the weak decay of the neutral kaons $K^0$, $\bar{K}^0$ in the standard formalism.
Consequently, a state at time $t$ is represented by
\begin{equation}
|\psi(t)\rangle =\sum_{j=1}^2 | j \rangle \langle j|\psi(t)\rangle,
\end{equation}
where $\{|1\rangle,|2\rangle\}$ correspond to $\{|K^0\rangle,|\bar{K^0}\rangle\}$ respectively. The evolution equation of the dynamics under consideration \eqref{Sch}, takes the form \cite{Lee}:
\begin{equation}\label{Dinamica en A}
\imath d_t\pmb{\psi}(t)=\mathbf{(M-\imath\Gamma)}\;\pmb{\psi}(t),
\end{equation}
where $\mathbf{M}$ and $\mathbf{\Gamma}$ are hermitian matrices
\begin{equation}
\mathbf{M-\imath\Gamma}=\left(\begin{matrix}
\mathrm{M}_{11}-\frac{i}{2}\,\Gamma_{11} & \mathrm{M}_{12}-\frac{i}{2}\,\Gamma_{12}\\
&\\
\mathrm{M}_{12}^{\ast}-\frac{i}{2}\,\Gamma_{12}^{\ast} & \mathrm{M}_{22}-\frac{i}{2}\,\Gamma_{22}
\end{matrix} \right).
\end{equation}
Clearly, the matrix $\boldsymbol{\Gamma}$ takes into account the decay width. To complete the physical description it is necessary to give the initial condition for the evolution.

All the information on the decay channels is contained in \eqref{Dinamica en A} as is clear from the matrix elements of $\mathbf{M-\imath\Gamma}$. The $CPT-$symmetry implies that $\mathrm{M}_{11}=\mathrm{M}_{22}$ and $\mathrm{\Gamma}_{11}=\mathrm{\Gamma}_{22}$, while if $T$ ($CP$) would be also a symmetry, then
$\mathrm{M}_{12}=\mathrm{M}_{12}^{\ast}$ and $\mathrm{\Gamma}_{12}=\mathrm{\Gamma}_{12}^{\ast}$, where $z^*$ is conjugate of the complex number $z$.

\subsubsection*{CP Violation}

After the crucial experiment \cite{Christenson}, it was clear that $CP$ symmetry was violated by weak interactions.
The eigenstates of $\mathbf{M-\imath\Gamma}$ expressed in the basis \{$|K^0\rangle,|\bar{K}^0\rangle$\} are now

\begin{equation}
\begin{split}
&|K_S\rangle = \dfrac{1}{\sqrt{2(1+|\epsilon|^2)}}
\Big[(1+\epsilon)|K^0\rangle + (1-\epsilon)|\bar{K}^0\rangle \Big]\\
&\\
&|K_L\rangle = \dfrac{1}{\sqrt{2(1+|\epsilon|^2)}}
\Big[(1+\epsilon)|K^0\rangle - (1-\epsilon)|\bar{K}^0\rangle\Big],
\end{split}
\end{equation}
where, as  usual, the indices $S,$ $L$ are realated to the decay times \textit{short, long} respectively and $\epsilon$ is a small parameter that 
measures the breaking of $CP$ symmetry and can be written \cite{Caruso:2011tu} in terms of the matrix elements of \textbf{H} ($\mathrm{H}_{ij}=\langle i|\mathbf{H}|j\rangle$) as
\begin{equation}
\epsilon = \frac{\sqrt{{\mathrm{H}_{12}}}-\sqrt{{\mathrm{H}_{21}}}} {\sqrt{{\mathrm{H}_{12}}}+\sqrt{{\mathrm{H}_{21}}}}.\label{Epsilon posta}
\end{equation}

It remains to present the time evolution of the solution of the quantum dynamical equation in the
basis $\{|K^0\rangle, |\bar{K^0}\rangle\}$
\begin{equation}  \label{wave}
|\psi(t)\rangle = \psi_1(t) \, |K^0\rangle + \; \psi_2(t) \,|\bar{K}^0\rangle.
\end{equation}

It is of interest to make explicit the
probability amplitudes

\begin{equation}\label{RRRwave}
\langle K^0|\psi(t)\rangle = \psi_1(t)
\end{equation}
and
\begin{equation} \label{RRwave}
\langle \bar{K}^0|\psi(t)\rangle = \psi_2(t).
\end{equation}

We chose a slightly different notation for the coordinates $\psi_j(t)$, given by \eqref{RRRwave} and \eqref{RRwave}, denoting the initial condition. We use $\psi_{ji}(t)$ for the $j-$component  of the solution $\pmb{\psi}(t)$ of \eqref{Dinamica en A} when the initial condition is $|\psi(0)\rangle=|\,i\,\rangle$, $i=1,2$, i.e. the system is prepared in state $|1\rangle =|K^0\rangle$ or $|2\rangle =|\bar{K}^0\rangle$, initally.

For the initial condition $|\psi(0)\rangle =|K^0\rangle$ the coordinates \eqref{RRRwave} and \eqref{RRwave} becomes
\begin{align}\label{qoordinates}
\psi_{11}(t) &= \frac{e^{k_St} + e^{k_Lt}}{2}, \nonumber\\
&\\
\psi_{21}(t) &= \left(\frac{1-\epsilon}{1+\epsilon}\right)\;\frac{e^{k_St} - e^{k_Lt}}{2}, \nonumber
\end{align}
where $k_S=-\Gamma_S/2-i m_S$ and  $k_L=-\Gamma_L/2-i m_L$ are the eigenvalues of $\mathbf{K}$.

Repeating the calculation for the initial condition $|\psi(0)\rangle =|\bar{K}^0\rangle$ the coordinates \eqref{RRRwave} and \eqref{RRwave} becomes
\begin{align}\label{qoordinatess}
\psi_{12}(t) &= \left(\frac{1+\epsilon}{1-\epsilon}\right)\;\frac{e^{k_St} - e^{k_Lt}}{2} \nonumber\\
&\\
\psi_{22}(t)& =\frac{e^{k_St} + e^{k_Lt}}{2}. \nonumber
\end{align}
this solutions are obtained directly from \eqref{qoordinates} interchanging the subindexes $j\longleftrightarrow i$ and $\epsilon\longrightarrow -\epsilon$.

The expressions \eqref{qoordinates} and \eqref{qoordinatess}, with these two initial conditions, will be useful to calculate transition probabilities, governed by the evolution operator $U(t)$. The quantum amplitude associate to the transition $|i\rangle\longmapsto|j\rangle$ is $A_{ji}(t)=\langle j| U(t) |i\rangle$ and also in this notation is given by
\begin{equation}
A_{ji}(t)=\psi_{ji}(t)
\end{equation}

Just to be ready to compare with the circuit signals, let us present the real part of the quantum probability amplitudes \eqref{qoordinates}, namely
\begin{align}\label{Rwave}
\operatorname{Re}\pmb{(}\psi_{11}(t)\pmb{)}&=
\frac{1}{2}\left[e^{-\Gamma_S t/2}\cos(m_S t) + e^{-\Gamma_L t/2}\cos(m_L t)\right]\nonumber\\
&\\
\operatorname{Re}\pmb{(}\psi_{21}(t)\pmb{)}&=\frac{f(\epsilon)}{2}\left[e^{-\Gamma_S t/2} \cos(m_S t) - e^{-\Gamma_L t/2}\cos(m_L t)\right],\nonumber
\end{align}
where $f(\epsilon)=\operatorname{Re}\left(\frac{1-\epsilon}{1+\epsilon}\right)$. 
Also real part of the quantum probability amplitudes \eqref{qoordinatess}, namely
\begin{align}\label{Rwave2} 
\operatorname{Re}\pmb{(}\psi_{12}(t)\pmb{)}&=
\frac{g(\epsilon)}{2}\left[e^{-\Gamma_S t/2}\cos(m_S t) - e^{-\Gamma_L t/2}\cos(m_L t)\right]\nonumber\\
&\\
\operatorname{Re}\pmb{(}\psi_{22}(t)\pmb{)}&=\frac{1}{2}\left[e^{-\Gamma_S t/2} \cos(m_S t) + e^{-\Gamma_L t/2}\cos(m_L t)\right],\nonumber
\end{align}
where $g(\epsilon)=\operatorname{Re}\left(\frac{1+\epsilon}{1-\epsilon}\right)$. 

The expressions \eqref{Rwave} and \eqref{Rwave2} are obtained from the condition $|\epsilon|\ll 1 $. The imaginary part will not be necessary due to the very good approximate validity of the Bedrosian theorem presented below.

\subsubsection*{Electric Networks}

Finally, let us introduce the electric networks of interest. As is well known, an electric network \cite{Bala,Carlin} includes a set of elements together with a given way of connections among them. These elements can be classified into five classes, namely: resistors $(R\,)$, capacitors $(C)$, inductances $(L)$, voltage generators $(v_s)$ and current generators $(i_s)$.
We are particularly interested in lumped element model circuits where voltage and current depend only upon time.

The corresponding dynamics of an electric network is defined by the appropriate use of the Kirchhoff rules that take care of the topology of the network. We restrict our analysis to passive networks, where the energy provided by an external source is non negative. The network has ports: pairs of terminals that allow to exchange energy with the surrounding and have a given voltage and current. One has the possibility of choosing the voltage or the current as the representative state variable of the excitation or the response of the network. We call $\pmb{V}$ the vector corresponding to the port voltage and $\pmb{I}$ the current.

A very important concept, relevant to our discussion, is that of \textit{reciprocity}. A network not connected to external energy sources is reciprocal iff considering two different terminals $\alpha$ $\neq$ $\beta$, the excitation in $\alpha$ gives rise to a response in $\beta$
that is invariant under the permutation $\alpha\longleftrightarrow\beta$.

In Ref.\cite{Caruso:2011tu}, an electric circuit that is equivalent to the system of neutral kaons (and eventually the other neutral mesons), in the sense previously introduced, was presented. As a result, the matrix elements of the effective Hamiltonian $\mathbf{H}$ were related, by means of a similitude transformation to those of the appropriate electric circuit.
In this way, the symmetries present in the kaon system and the corresponding tests of validity, have a unique reflection in the electric circuit.

\begin{figure}[!ht]
\centering
\includegraphics[height=4cm]{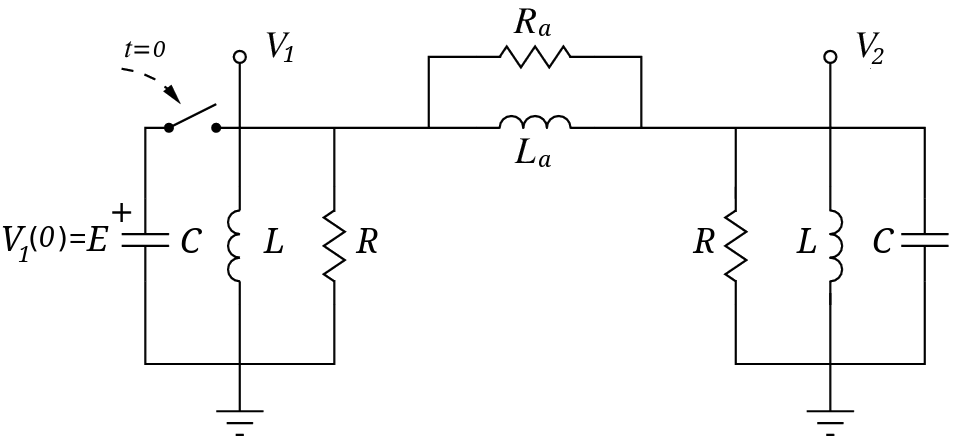}
\centering
\hspace*{0.5cm}
\includegraphics[height=5cm,trim=1cm 0.5cm 1cm 0.5cm, clip=true]{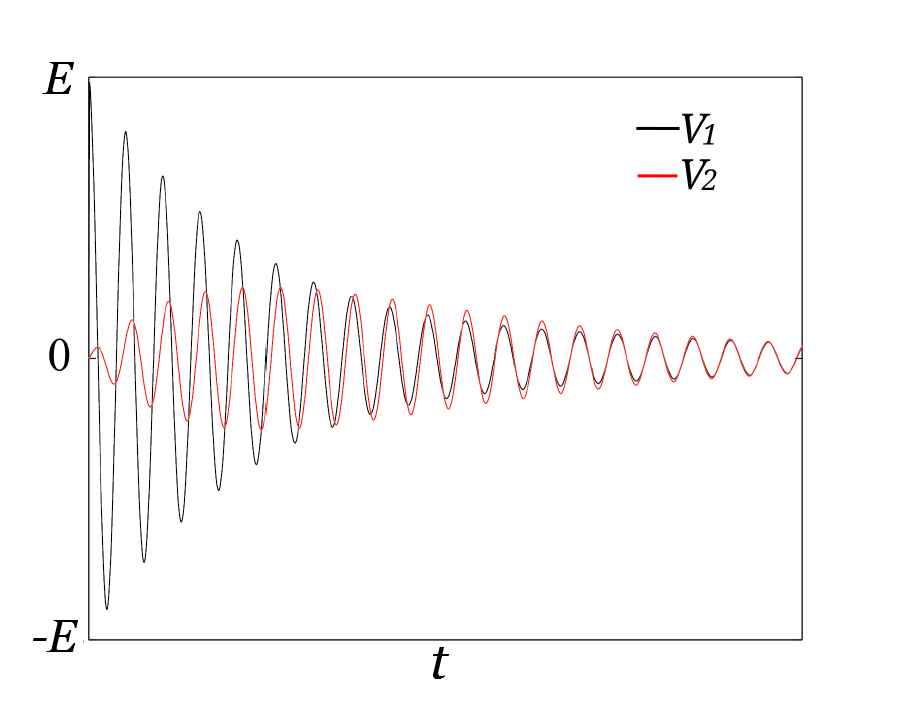}
\caption{\footnotesize{$CP$ conserving electric network equivalent. LEFT: Schematic of the ideal circuit. RIGHT: Simulation of port time response to an initial condition in left $C$.
\label{figcircuitosimple}}}
\end{figure}

The analysis of the time evolution of electric circuits results in a system of linear differential equations with constant coefficients as Eq.\eqref{A B}. The \textit{synthesis} \cite{Bala} of all electric networks in a given family was analyzed in Ref.\cite{Caruso:2011tu} and ends, in the case of exact $CP$ symmetry, with the simple circuit in Fig. \ref{figcircuitosimple}. This circuit, due to the presence of a loop of inductances, has two proper frequencies \cite{Bala}.

Notice that the equations that govern the circuit in Fig. \ref{figcircuitosimple} will be given as a particular case of the more general ones stated below when the non-reciprocal elements are included.

The next step is to find a modified circuit in order to take into account the $CP$ violation experimentally present in the kaon system.
A brief review of our previous analysis shows that the only way of breaking the $CP$ symmetry is  the interaction network being  \textit{non-reciprocal} \cite{Caruso:2011tu}.

Due to the fact that any combination of $\{R, L, C\}$ elements provides
a reciprocal network \cite{Bala}, the introduction of some new kind
of component is unavoidable. A gyrator, which is a passive element of
two ports, does the job \cite{Tellegen}.

As a consequence of the introduction of a gyrator of conductance $g$ in a circuit, the
admittance (or impedance) matrix is not symmetric anymore.

\begin{figure}[!ht]
\centering
\hspace*{0.1cm}\includegraphics[height=4cm]{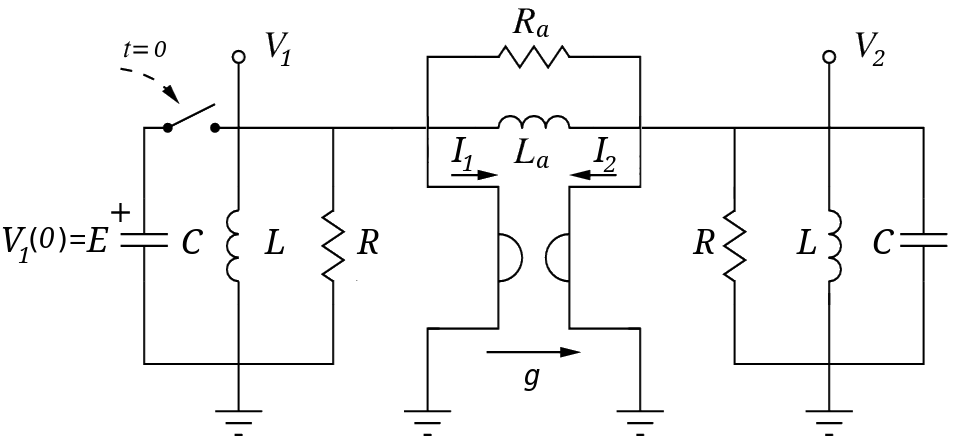}
\centering
\includegraphics[height=5cm, trim=1cm 0.5cm 1cm 0.5cm]{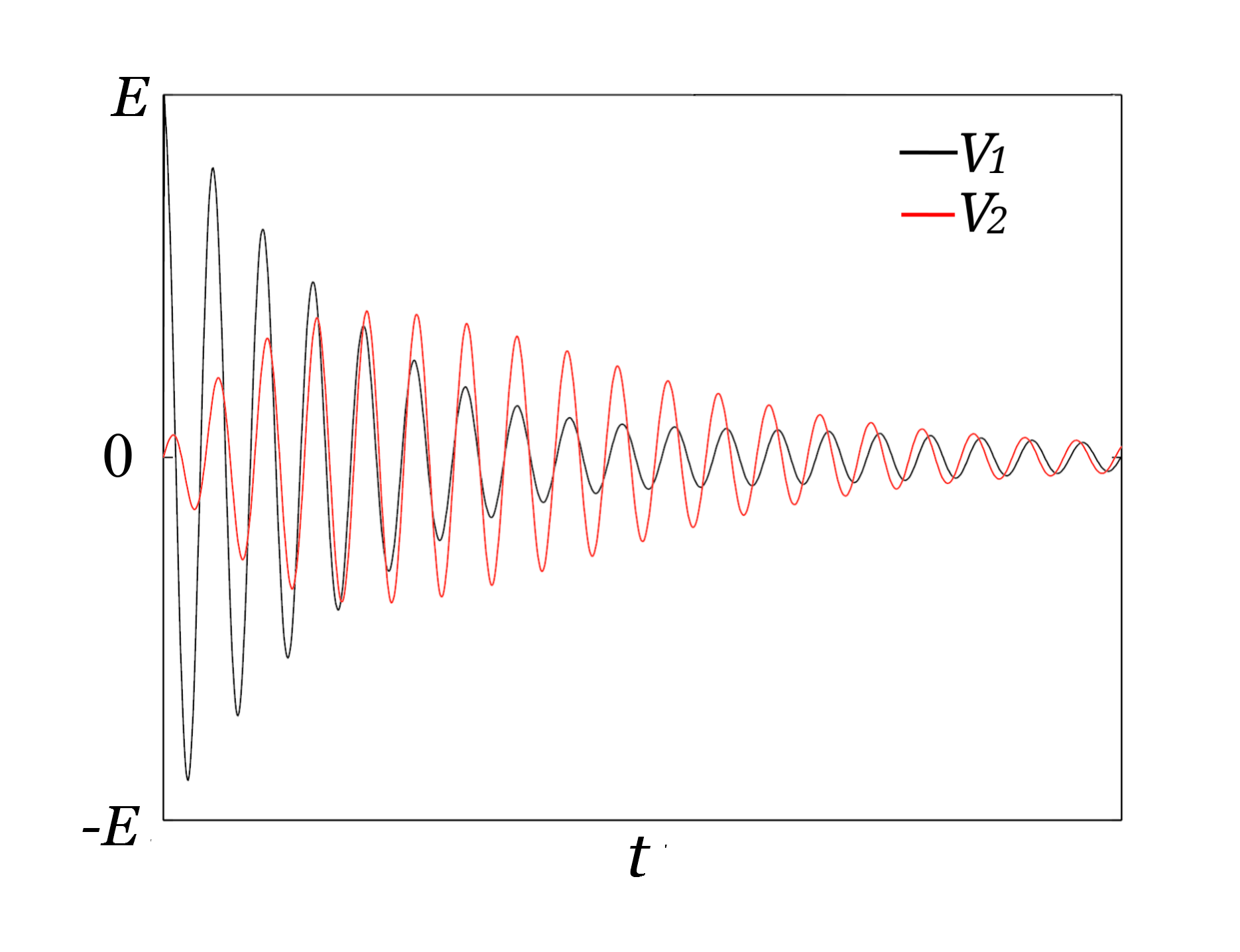}
\caption{\footnotesize{Non-reciprocal electric circuit. LEFT: Schematic of the ideal circuit. RIGHT: Simulation of port time response to an initial condition in left $C$. \label{Girador}}}
\end{figure}

Under the hypothesis that the non reciprocity is very small because the $CP$ violation is measured by a parameter of the order $|\epsilon| \sim 10^{-3}$, one has to deal with a small perturbation on the initial reciprocal circuit and consequently there is not a measurable change of the proper frequencies of the symmetric system. When the gyrator is included in the coupling between the original oscillators, the circuit it that shown in Fig. \ref{Girador}

Now the equations relating current and voltage through the gyrator read
\begin{equation}
I_1=gV_{2} \: ,\quad I_2=-gV_{1}.
\end{equation}
We consider for the moment node $1$, because for node $2$ the situation is entirely similar.  Current conservation implies $I_{1}+I_{L}+I_{C}+I_{R}+I_{R_a}+I_{L_a}=0$. From here one directly obtains
\begin{equation*}
\ddot{V}_1+(\gamma+\gamma_{a})\dot{V}_1+(\omega
_{0}^{2}+\omega_{a}^{2})V_{1}+(\gamma_{g}-\gamma_{a})\dot{V}_2-\omega_{a}^{2}V_{2}=0,
\end{equation*}
where the parameters are defined through
\begin{equation*}
\gamma=\frac{1}{RC},\;  \gamma_{a}=\frac{1}{R_{a}C} ,\; \gamma_{g}=\frac{g}{C},\;\omega_{0}^{2}=\frac{1}{LC}, \; \omega_{a}^{2}=\frac{1}{L_{a}C}.
\end{equation*}
In the same way, for the node $2$ one gets, merely exchanging $1$ by $2$ and the gyrator sign
\begin{equation*}
\ddot{V}_2+(\gamma+\gamma_{a})\dot{V}_{2}+(\omega
_{0}^{2}+\omega_{a}^{2})V_{2}-(\gamma_{g}+\gamma_{a})\dot{V}_1-\omega_{a}^{2}V_{1}=0,
 \end{equation*}
where we have omitted the time dependence for simplicity.

The last two differential classical equations can be summarized as a system of differential equations
\begin{equation}\label{sist clas}
\ddot{\pmb{V}}+\mathbf{A}\dot{\pmb{V}}+\mathbf{B}\pmb{V}=\pmb{0},
\end{equation}
where $\pmb{V}$, $\mathbf{A}$ and $\mathbf{B}$ are given by 
\begin{equation}\label{Vec}
\pmb{V}=(V_1,V_2)^\intercal
\end{equation}
\begin{align} \label{ABmatrices}
\mathbf{A}=\left(\begin{matrix}
\hspace*{0.42cm}\gamma + \gamma_a && \gamma_g-\gamma_a\\
              -\gamma_g-\gamma_a	 &\;&\hspace*{.17cm} \gamma + \gamma_a
\end{matrix}\right)\nonumber\\
&\\
\mathbf{B}=\left(\begin{matrix}
\omega_0^2 + \omega_a^2 & \;&-\omega_a^2\\
              -\omega_a^2 &\;&  \omega_0^2 + \omega_a^2
\end{matrix}\right).\nonumber
\end{align}
The system \eqref{sist clas}, with \eqref{ABmatrices}, corresponds to a \textit{non-normal} system of differential equations because $[\mathbf{A},\mathbf{B}]\neq\mathbf{0}$ \cite{Politzer}.
However it is a special one, because the characteristic polynomial associated $\mathbf{A}$ is equal (at order $\gamma_g^2$) to the case of $\gamma_g = 0$.

Let us see this as follows from $\mathbf{A}_0 = \mathbf{A}|_{\gamma_g=0}$, then the difference of the characteristic polynomial of $\mathbf{A}$ and $\mathbf{A}_0$ is equal to $\gamma_g^2$. Therefore the eigenvalues of $\mathbf{A}$ are almost
equal to the $\mathbf{A}_0$ at order $\gamma_g^2$.

We will demonstrate that there is an explicit correspondence between the solutions of \eqref{Dinamica en A} and \eqref{sist clas} given by
\begin{equation}
\operatorname{Re}\pmb{(\psi)}\longleftrightarrow \pmb{V}.
\end{equation}

Introducing the eigenvalues of $\mathbf{A}_0$ and $\mathbf{B}$

\begin{eqnarray} \label{omegacero0}
&\Gamma_{+}=\gamma \qquad &\Gamma_{-}=\gamma+2\gamma_{a}\nonumber\\
&\\
&\omega_{+}^{2}=\omega_{0}^{2}\qquad &\omega_{-}^{2}=\omega_{0}^{2}+2\omega_{a}^{2}.\nonumber
\end{eqnarray}
Referring now to the parameters related to the damping, one defines
\begin{align}\label{Deltas}
\Delta \omega&=\omega_{+}-\omega_-\nonumber\\
&\\
\Delta\Gamma&=\tfrac{1}{2}(\Gamma_{+}-\Gamma_{-})\nonumber
\end{align}
We considered that
\begin{align}
\Gamma_+\leq\Gamma_- &\ll 2\omega_\pm\nonumber\\
&\nonumber\\
2\gamma_g &< \Gamma_\pm\\
&\nonumber\\
0\simeq|\Delta \omega| &\ll|\Delta\Gamma|.\nonumber
\end{align}
Under the approximations mentioned above, the modes of the damped coupled equations \eqref{sist clas} are not changed. The term \textit{modes} here refers simply to the roots of the characteristic polynomial associated the system \eqref{sist clas} with \eqref{ABmatrices}.

As we did in section \ref{cap1}, we chose a notation for the coordinates $V_j(t)$, given by \eqref{Vec}, denoting the initial condition. We use $V_{ji}(t)$ for the $j-$component of the solution $\pmb{V}(t)$ of \eqref{sist clas} when the initial condition is the state corresponds to the excitation of the  $i-$node only.

We associate the state $|K^0\rangle$ with the left oscillator in Fig. \ref{Girador} at an initial time. Consequently, the case that $|\psi(0)\rangle=|K^0\rangle$ corresponds to the excitation of the node $1$ only, i.e., an initial condition $V_{11}(0)=E$, $V_{21}(0)= 0$ and $\dot{V}_{11}(0)=0=\dot{V}_{21}(0)$, therefore the solutions result in
\begin{align}\label{v}
V_{11}(t)&=\frac{E}{2}\;\;\left[ e^{-\Gamma_+/2 \,t}cos (\omega_+\,t)+e^{-\Gamma_-/2 \,t}cos (\omega_-\,t)\right]\nonumber\\
&\\
V_{21}(t)&=\frac{\mu\,E}{2}\left[e^{-\Gamma_+/2 \,t}cos (\omega_+\,t) -e^{-\Gamma_-/2 \,t}cos (\omega_-\,t)\right],\nonumber
\end{align}
where $\mu=\left(1+\frac{\gamma_g}{\gamma_a}\right)$.

There is an explicit correspondence between the classical and quantum coordinates only if there is an
identification
\begin{align}\label{xi}
\operatorname{Re}\left(\frac{1-\epsilon}{1+\epsilon}\right)&\longleftrightarrow \left(1+\frac{\gamma_g}{\gamma_a}\right)
\end{align}
\begin{align}
\left(\Gamma_L,\Gamma_S,m_L,m_S\right)&\longleftrightarrow \left(\Gamma_+,\Gamma_-,\omega_+,\omega_-\right)
\end{align}
From \eqref{xi}, using 	$|\epsilon|\ll 1$ and $arg(\epsilon)=\pi/4$  (\textit{or} $5\pi/4$) \cite{Lee} we have
\begin{align}\label{epsxi}
|\epsilon|&\longleftrightarrow\frac{\gamma_g}{\sqrt{2}\gamma_a}.
\end{align}
The right side of the correspondence \eqref{epsxi} is the \textit{classical} quantity associated $|\epsilon|$, denoted by
\begin{equation}\label{epsilon clasico}
\xi=\frac{\gamma_g}{\sqrt{2}\gamma_a}.
\end{equation}
This clearly allows the announced identification of the $CP$ violation parameter with the circuit parameters. Also shows that in the implementation of the circuit one faces a compromise between the coupling of the separate initial oscillators and the effect of the gyrator.\\

Moreover, the case that $|\psi(0)\rangle = |\bar{K}^0\rangle$ corresponds to the excitation of the node $2$ only, i.e., an initial condition $V_{22}(0)= E$, $V_{12}(0) = 0$ and $\dot{V}_{12}(0) = 0 = \dot{V}_{22}(0)$. As we did in section \ref{cap1}, the solutions are obtained directly from \eqref{v} interchanging the subindexes $j\longleftrightarrow i$ and $g\longrightarrow -g$, therefore the solutions result in
\begin{align}\label{vb}
V_{12}(t)&=\frac{\nu\,E}{2}\left[ e^{-\Gamma_+/2 \,t}cos (\omega_+\,t)-e^{-\Gamma_-/2 \,t}cos (\omega_-\,t)\right]\nonumber\\
&\\
V_{22}(t)&=\frac{E}{2}\;\;\left[e^{-\Gamma_+/2 \,t}cos (\omega_+\,t) +e^{-\Gamma_-/2 \,t}cos (\omega_-\,t)\right],\nonumber
\end{align}
where $\nu=\left(1-\frac{\gamma_g}{\gamma_a}\right)$.
One should also notice that the non-reciprocity introduced by means of the gyrator is present only in $\mathbf{A}$ \eqref{ABmatrices}, while the $CP$ violation is present also in $|\mathbf{K}|^2:=\mathbf{K}_r^2+\mathbf{K}_i^2$ \eqref{soe}. However, if the quantum system is prepared with an initial condition $|\psi(0)\rangle = |1\rangle$, the port voltage $V_{11}$ and $V_{21}$ \eqref{v} are entirely similar to the real part of  $\psi_{11}$ and  $\psi_{21}$ given by \eqref{qoordinates}. On the other hand if the quantum system is prepared with an initial condition $|\psi(0)\rangle = |2\rangle$, the port voltage $V_{12}$ and $V_{22}$ \eqref{vb} are entirely similar to the real part of  $\psi_{12}$ and $\psi_{22}$ given by \eqref{qoordinatess}.
In summary, we have presented an explicit correspondence between the quantum and classical coordinates	
\begin{align}
&\operatorname{Re}\pmb{(}\psi_{ji}(t)\pmb{)}\longleftrightarrow V_{ji}(t)
\end{align}	
and we are ready to implement this correspondence experimentally.

\section{Building Observables}

This section is devoted to the presentation of the physical observables in both systems, the neutral kaons and the electrical network.

The analysis is simplified when the concept of {\it analytic signal} \cite{TT} is introduced. Let us consider the voltage signal $\pmb{V}(t)$.  It can be expressed in terms of the Fourier representation
\begin{equation}
\pmb{V}(t) = \int_{-\infty}^{\infty}\,\pmb{v}(\omega)\,e^{-2 \pi \imath \omega t}\,d\omega .
\end{equation}
If the signal is real, one has $\pmb{v}(- \omega) = \pmb{v}^{\ast}(\omega)$, that means that the positive frequency already
contains all the information. Given a real
signal $\pmb{V}(t)$, the analytic signal is introduced through
\begin{equation}
\pmb{V_a}(t) = 2 \int_{0}^{\infty}\,\pmb{v}(\omega)\,e^{-2 \pi \imath \omega t}\,d\omega,
\end{equation}
clearly we have
\begin{equation}\label{s-s_a(1)}
\pmb{V}(t) = \operatorname{Re}\pmb{(} \pmb{V_a}(t)\pmb{)}.
\end{equation}
\\
Consequently, $\pmb{V_a}(t)$ is a complex signal having the actual signal as the real
part and the Hilbert transform of the signal as the imaginary component, namely;
\begin{equation}\label{as}
\pmb{V_a}(t) = \pmb{V}(t) + \imath\, H\pmb{(}\pmb{V}(t)\pmb{)},
\end{equation}
where $H$ is the Hilbert transform defined as
\begin{align}
H\pmb{(}\pmb{V}(t)\pmb{)} & = \frac{1}{\pi}\, \mbox{{\Large \textit{P}}}  \int_{-\infty}^{\infty}\,\frac{\pmb{V}(t')}{t-t^{\prime}}\,dt^{\prime},
\end{align}
where {\large\textit{P}} denotes the Cauchy principal value. The Hilbert transform relates the real and imaginary parts of the analytic signal:
\begin{align}
\operatorname{Im}\pmb{(}\pmb{V_a}(t)\pmb{)} & = \frac{1}{\pi}\, \mbox{{\Large \textit{P}}} \int_{-\infty}^{\infty}\,\frac{\operatorname{Re}\pmb{(}\pmb{V_a}(t)\pmb{)}}{t^{\prime} - t}\,dt^{\prime} \\
&\nonumber\\
\operatorname{Re}\pmb{(}\pmb{V_a}(t)\pmb{)} & = \frac{1}{\pi}\, \mbox{{\Large \textit{P}}} \int_{-\infty}^{\infty}\,\frac{\operatorname{Im}\pmb{(}\pmb{V_a}(t)\pmb{)}}{t-t^{\prime}}\,dt^{\prime} = \pmb{V}(t). \label{s-s_a(2)}
\end{align}
Notice that the Hilbert transform $H\pmb{(}\pmb{V}(t)\pmb{)}$ satisfies the same differential equation as $\pmb{V}(t)$. The use of the analytic signal allows a closer contact with quantum-mechanical descriptions.

In the process of comparison of observables in our systems, we have to take into account the fact that the resulting signal in both cases
is composed (see for example Eqs. \eqref{Rwave}, \eqref{v}) by a rapidly varying part ($\sim\cos(\omega_{\pm} t)$) modulated by a slowly varying term ($\sim e^{\Gamma_{\pm} t}$).
This particular situation allows the use of the Bedrosian theorem \cite{Bedrosian} that states\vspace*{0.3cm}

{\it Let $f$ and $g$ $\in L^2(\mathbb{R})$. Suppose
that the Fourier transform of $f(x)$, $F(\omega)$, vanishes for $|\omega| > a$, with $a \in\mathbb{R}^+$
and the Fourier transform of $g(x)$, $G(\omega)$, vanishes for $|\omega| < a$; then
$H\pmb{(}f(x)\,g(x)\pmb{)} = f(x)\,H\pmb{(}g(x)\pmb{)}$.}
\vspace*{0.3cm}

This theorem is, with very good precision, valid in our case, due to fact that the spectra of the signal have very separate frequencies.
As a consequence of the validity of the theorem, one can consider only the real part of the solution and from it to construct, via the Hilbert transform, the corresponding imaginary part. We said that $\operatorname{Re}\pmb{(}\psi_i(t)\pmb{)}\longleftrightarrow V_i(t)$, for $i=1,2$; from the last theorem we complete the sentence as
$\operatorname{Im}\pmb{(}\psi_i(t)\pmb{)}\longleftrightarrow H\pmb{(}V_i(t)\pmb{)}$. Therefore the vector of quantum coordinates $\pmb{\psi}=(\psi_1,\psi_2)^\intercal$ of \eqref{qoordinates} is related to the analytic signal of $\pmb{V}=(V_1,V_2)^\intercal$, $\pmb{V_a}$ according to \eqref{as}, as
\begin{equation}\label{EQUI}
\pmb{\psi}\longleftrightarrow \pmb{V_a}.
\end{equation}
The quantum amplitudes of probabilities are given by the analytic signals of the real parts, equivalent of port voltages. As a corollary the probabilities, defined as the square of the quantum amplitudes, are given by the \textit{envelope} of these classical signals. 

If we define an operator $\mathcal{A}$ such that returns the analytical signal, we have $\mathcal{A}(\pmb{V})=\pmb{V_a}$, from \eqref{EQUI}
\begin{equation}\label{ampl analsign}
A_{ji}(t)\longleftrightarrow \mathcal{A}\pmb{(}V_{ji}(t)\pmb{)}.
\end{equation}

\subsection{Neutral Kaons}

The physical magnitudes of the kaon system  that are of interest for the comparison with the equivalent electrical network are related with time-dependent probabilities \cite{Reiser:2012fh}. These  are expressed as 
\begin{equation}
P_{ji}(t)=|\langle j |U(t) |i\rangle|^2,
\end{equation}
where $|i\rangle$ and $|j\rangle$ are the initial and final states, respectively. In particular, $i,j=1,2$ and again the states $\{|1\rangle,|2\rangle\}$ correspond to $\{|K^0\rangle,|\bar{K^0}\rangle\}$ respectively. These quantities are interpreted as conditional probabilities (transition probabilities) to start in the state $|i\rangle$ and evolve at state $|j\rangle $ at time $t$.

It is clear that the validity of $CPT-$symmetry implies that
\begin{equation}
P_{11}(t)=P_{22}(t),
\end{equation}
while the $CP$, or $T$, violation manifest itself by the inequality
\begin{equation}\label{PD}
P_{21}(t)\neq P_{12}(t),
\end{equation}
showing the non-reciprocity of the kaon system.

Any one of the probabilities mentioned above are obtained from the corresponding wave function. They are all of the type presented
in Eq. \eqref{wave} and due to the validity of the Bedrosian theorem, they can be expressed in terms of only the real (or the imaginary) part of
the wave function. It is worth to remark that this possibility, that seems to indicate that in quantum mechanics the imaginary (or the real) part is almost superfluous, is only a particularity of systems such as the kaon one, where the spectrum of frequencies involved defines two very separated regimes (the overlapping is negligible). \\
\\
The probabilities $P_{ji}(t)$ are equal to $|\psi_{ji}(t)|^2$ given by \eqref{qoordinates} and \eqref{qoordinatess}, according to the initial condition $i=1,2$.
The explicit expressions for the probabilities are
\begin{align}
P_{11}(t) =&\frac{1}{4}\,\bigg[e^{-\Gamma_S\,t} + e^{-\Gamma_L\,t}+ 2\,e^{-(\Gamma_S+\Gamma_L)\,t/2}\,\cos(\Delta m\,t)\bigg], \label{P1} \\[1em]
P_{21}(t) =&\frac{1}{4}\,|1-2\epsilon|^2\,\bigg[e^{-\Gamma_S\,t} + e^{-\Gamma_L\,t} - 2\,e^{-(\Gamma_S+\Gamma_L)\,t/2}\,\cos(\Delta m\,t)\bigg], \label{P2}  \\[1em]
P_{12}(t) =&\frac{1}{4}\,|1+2\epsilon|^2\,\bigg[e^{-\Gamma_S\,t} + e^{-\Gamma_L\,t} - 2\,e^{-(\Gamma_S+ \Gamma_L)\,t/2}\,\cos(\Delta m\,t)\bigg],\label{P3}\\[1em]
P_{22}(t) =&\frac{1}{4}\,\bigg[e^{-\Gamma_S\,t} + e^{-\Gamma_L\,t} + 2\,e^{-(\Gamma_S+\Gamma_L)\,t/2}\,\cos(\Delta m\,t)\bigg],\label{P4}
\end{align}
where $\Delta m = m_L - m_S$ and $|1\pm 2\epsilon|^2\simeq 1\pm4\operatorname{Re}(\epsilon)$, under $|\epsilon|\ll 1$.

The probabilities $P_{ji}$ are associated to $|\mathcal{A}(V_{ji})|^2$ \eqref{ampl analsign}. In particular, $P_{j1}$ is related to $|\mathcal{A}(V_{j1})|^2$ given by \eqref{v} for $j=1,2$, respectively. And $P_{j2}$ is related  to $|\mathcal{A}(V_{j2})|^2$ given by \eqref{vb} for $j=1,2$, respectively. In summary we have

\begin{align}
|\mathcal{A}\pmb{(}V_{11}(t)\pmb{)}|^2 =&  \frac{E^2}{4}\,\bigg[e^{-\Gamma_-\,t} + e^{-\Gamma_+\,t} + 2\,e^{-(\Gamma_- +\Gamma_+)\,t/2}\,\cos(\Delta \omega\, t)\bigg],\label{PP1} \\[1em]
|\mathcal{A}\pmb{(}V_{21}(t)\pmb{)}|^2  =& \frac{\mu^2\,E^2}{4}\,\bigg[e^{-\Gamma_-\,t} + e^{-\Gamma_+\,t}- 2\,e^{-(\Gamma_- +\Gamma_+)\,t/2}\,\cos(\Delta \omega\,t)\bigg], \label{PP2}   \\[1em]
|\mathcal{A}\pmb{(}V_{12}(t)\pmb{)}|^2  =& \frac{\nu^2\,E^2}{4}\,\bigg[e^{-\Gamma_-\,t} + e^{-\Gamma_+\,t}- 2\,e^{-(\Gamma_- + \Gamma_+)\,t/2}\,\cos(\Delta \omega\,t)\bigg], \label{PP3}\\[1em]
|\mathcal{A}\pmb{(}V_{22}(t)\pmb{)}|^2  =& \frac{E^2}{4}\,\bigg[ e^{-\Gamma_-\,t} + e^{-\Gamma_+\,t} + 2\,e^{-(\Gamma_- +\Gamma_+)\,t/2}\,\cos(\Delta \omega\,t)\bigg]. \label{PP4}
\end{align}

\subsection{Electric Network}

The previous discussion of the particularities of the quantum system under consideration has its reflection in the classical system. In fact, it is not necessary to consider, as was mentioned before, two identical circuits with different initial conditions in order to maintain the complex character of the quantum equivalent system.

The analysis of the classical signal, in our case the electric voltage, clearly shows that in making the comparison of observables,
corresponding to the quantum and the classical systems, the analytic signal is obtained from the measurement of the voltage and
can be put in direct connection with the wave function of kaons.

\section{Simulation and Experimental Results}

\subsection{Resonant Network}

The $CP-$conserving electric circuit consists of two identical resonant $LCR$  networks, coupled by a parallel $L_aR_a$ impedance. LTSpice \cite{spice}  simulations were used in the design process to solve several implementation tradeoffs and to calculate port output as in Fig. \ref{figcircuitosimple}. The quotient $\frac{L_a}{L}$ determines the relationship between fast and slow dynamics. This relation cannot be arbitrarily chosen. For instance, setting too fast an oscillation frequency increases energy dissipation, thus making the phenomenon almost invisible due to excessive damping.

Two identical inductors were made for $L$, while $L_a$ was selected one order of magnitude higher. A value of $L_a/L=25$ was chosen, which allows adequate filtering of the individual dynamics, and still allows application of the Bedrosian theorem in the calculation of Hilbert transforms. Inductors were made using copper wire wound on ferrite nuclei, resulting in values of $L=0.7$ mHy (pot core, $Q=70$) and $L_a=18$ mHy (toroidal, Q=150). The wire used to implement $L$ introduces a parasitic resistance (in series with each inductor) of approximately 2 $\Omega$. These are critical in the experimental realization, being responsible for the resonance attenuation.  Of course, the series resistance of $L_a$ is even higher. These unavoidable resistances make the actual circuit differ from the ideal case. In fact, in the real circuit the oscillations vanish after a few milliseconds. This establishes an important difference between the actual circuit behavior and its kaon counterpart. In the real case, there appears a third proper frequency due to these unavoidable resistances. This new mode vanishes almost immediately and does not obscure the analysis. The parasitic resistances were included in all the Spice simulations performed.

For $C$, polyester capacitors of $0.1$ $\mu$F were selected, in order to achieve proper resonant frequencies. Obviously, in the experimental setup imbalances exist between the $LCR$ subnetworks. Therefore, small-value capacitors were added in parallel with $C$, to allow experimental tuning of the individual resonance frequencies. Additionally, it must be noted that a limitation exists in the values that $R$ can take for simulations and implementation, as they are connected in parallel with the loss resistances of $C$, which cannot be modified. 

Chosen values result, from \eqref{omegacero0} and \eqref{Deltas}, in the circuit parameters presented in table \ref{tab:table1}.

\begin{table}[ht!]
  \begin{center}
    \caption{Circuit parameters.}
    \label{tab:table1}
    \begin{tabular}{l|c|r}
      $f_{+}$   &   $\frac{1}{2 \pi} \sqrt {\frac {1}{L C} }$   &   $19.02\, \mathrm{KHz}$ \\
      \hline
      $f_{-}$   &   $\frac{1}{2 \pi} \sqrt {\frac {1}{L C} + \frac{2}{L_a C} }$   &   $19.74\, \mathrm{KHz}$  \\
      \hline
      $\Delta f$   &   $f_{+} - f_{-}$   &   $0.72\, \mathrm{KHz}$   \\
      \hline
      $\bar{f} $   &   $\frac{1}{2} (f_{+} + f_{-})$   &   $19.38\, \mathrm{KHz}$   \\
    \end{tabular}
  \end{center}
\end{table}

The relation $\bar f/\Delta f$ is given by $L_a/L$. It fixes the relation between the two distinguishable frequencies in port voltages. The component values chosen result in $\gamma_a^{-1} = 0.3$ ms and $\gamma_g^{-1} = 1$ ms, while $\gamma^{-1} = 0.23$ ms, which is imposed  by the parasitic resistances.

Port voltages were acquired using an Agilent MSO-X 2024A 200 MHz osciloscope in averaging mode, and then processed using Matlab\textregistered. This processing involves computing the discrete Hilbert transform of the acquired data and multiplying it by its complex conjugate, in order to obtain the squared envelope signal. {A brief comment about Matlab: The command $\mathtt{hilbert(x)}$ returns the complete analytic signal of $\mathtt{x}$.

The circuit was also simulated using LTSpice, including all the parasitic resistances. Simulated port voltages were processed using the same Matlab algorithm.

\subsection{Gyrator implementation issues}

The gyrator is a hypothetical circuit element that is passive and lossless. It does not exist as a physical element. However, it is certainly possible to build an active circuit which behaves as a gyrator nearby a given operational point. Following \cite{Gira}, such a device was built using a dual operational transconductance amplifier (Texas Instruments LM13700). This device features excellent matching between amplifiers.

Generally speaking, an operational transconductance amplifier (OTA) is a device that acts as a voltage controlled current source. It has the convenient feature of requiring the modification of a single parameter (the amplifier bias current from an external source) to change its transconductance value. This is accomplished by using an external resistor connected to a dc source.

In the gyrator implementation, two OTAs with equal transconductance values are interconnected (with opposite polarities), one in the forward and other in the backwards direction, as shown in Fig. \ref{Impl1} (left). This arrangement effectively behaves as a gyrator, featuring $g$ in the forward path and $-g$ in the backwards path, as was experimentally verified in \cite{Gira}. The parameter $g$ results from the transconductance value of the amplifiers, which must strictly match.

One of the main drawbacks of the resulting circuit is that a modification of $g$ requires the simultaneous change of two precision resistors (one for each OTA). For this reason the circuit was calibrated for a single, fixed value $g=0.1 mS$. The calibration procedure involves the trimming of an external input resistor (not shown) in each amplifier, in order to ensure matching $g$ values in both paths. This value was used as a starting point for the  design.

Although the circuit has the desired behavior for this application, it differs from the ideal gyrator in several aspects. On the one hand, its dynamic range, bandwidth and rise times are limited by the characteristics of the transconductance amplifiers. This issue was minimized restricting operating frequencies to a few KHz and ensuring small signal amplitudes. On the other hand, like in any operational amplifier, it presents nonlinearities that could affect signal amplitudes, eventually resulting in distortion. In order to improve linearity, the LM13700 internal output buffers were not used. Instead, external noninverting buffers (followers), based on a CA324 dual operational amplifier (not shown in the figure) were included in the feedback path. In this way, nonlinearities issues are negligible for the operational conditions devised.  This allows modelling of the gyrator in the simulations as two ideal voltage-dependent current sources.

\subsection{Setting the initial conditions}

In order to replicate the initial conditions used in simulations, and display the desired circuit behavior as a steady image in the oscilloscope, two basic requirements must be met: a) an initial charge in one of the capacitors must be ensured while the other initial conditions of the circuit are null (the other capacitor is discharged and there is no current in all inductors) and b) this situation must be repetitive. Therefore, additional circuitry was included to disconnect one of the capacitors from the rest of the circuit, charging it to a known state ($V_1(0)=V_{cc}$ in Fig. \ref{Impl1}), and reconnecting once the transient has vanished. This was implemented with a pair of MOSFET power transistors (IRFD9110 N channel and IRFD420 P channel), driven by a square wave provided by an Agilent 8648C signal generator. During operation, the ON-state of the P channel transistor (1 $\Omega$) appears in series with left $C$, increasing dissipation in the resonant circuit .

The complete experimental circuit is shown in Fig. \ref{Impl1} (left). The layout permits enabling/disabling gyrator operation, as well as inversion of the gyrator ports.
\begin{figure*}[!h]
\centering
\includegraphics[width=4.4in]{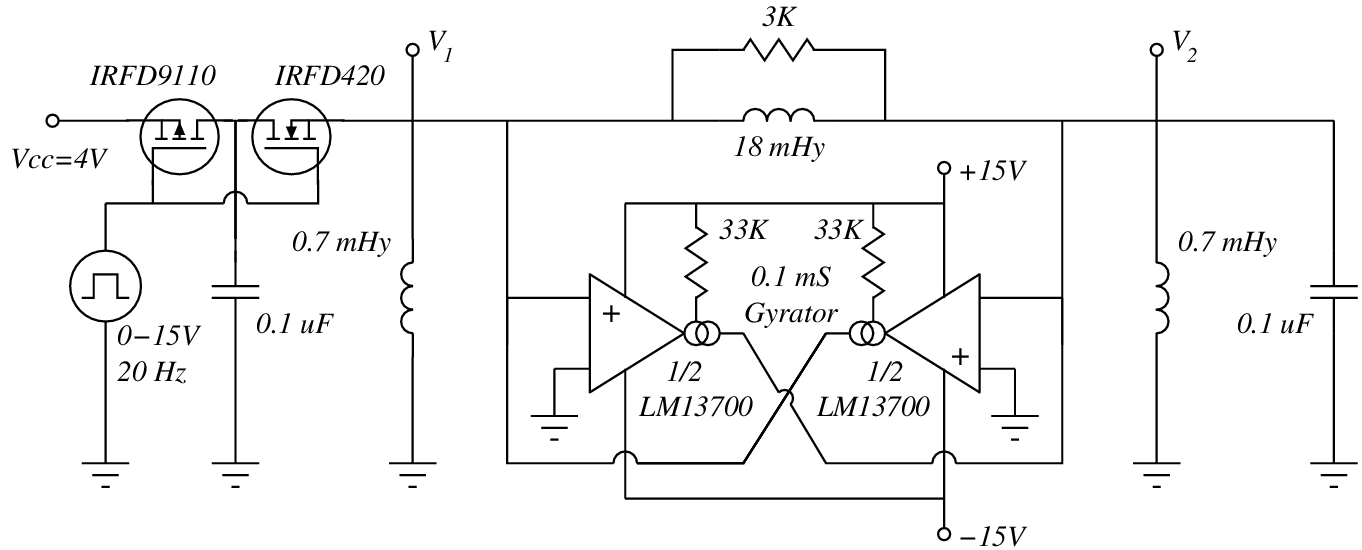}
\includegraphics[width=2.4in]{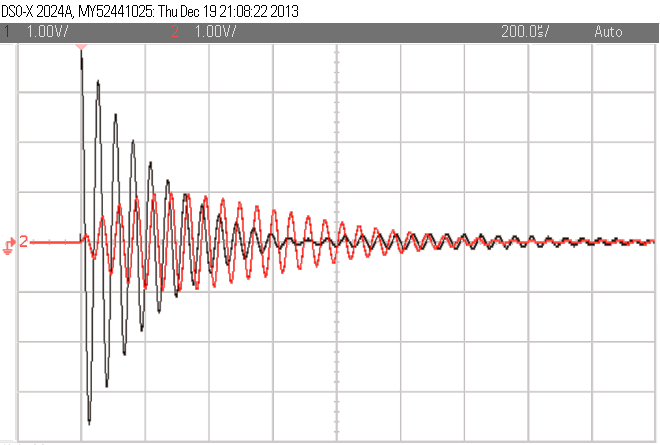}
\caption{\footnotesize{Experimental Implementation.
LEFT: Schematic of the real circuit. RIGHT: Measured time response of the real circuit.\label{Impl1}}}
\end{figure*}

\newpage
\subsection{Final results and comments}

In Fig. \ref{Impl2} and \ref{Impl3} simulation and experimental results are presented respectively. There one can easily observe the following: When $g=0$, namely, when no gyrator is included, both probabilities $P_{11}$ and $P_{21}$ from \eqref{P1} and \eqref{P2} are correlated to $|\mathcal{A}(V_{11})|^2$ and $|\mathcal{A}(V_{21})|^2$ from \eqref{PP1} and \eqref{PP2}, having exactly the same asymptotic behavior. When the gyrator is acting, the second figures show the probabilities correlated to \eqref{PP1} and \eqref{PP2} having, as expected, a different asymptotic behavior, measured by $g$. The third figures include the corresponding probabilities \eqref{P3} and \eqref{P4}, correlated to \eqref{PP3} and \eqref{PP4}, with an entirely similar behavior to the previous ones. Finally the fourth figures show clearly the non-reciprocity effect present in \eqref{PP2} and \eqref{PP3} and equivalent to the $CP$ or $T$ violation established in \eqref{PD}.

\begin{figure}[!h]
\centerline{\includegraphics[totalheight=5.5cm]{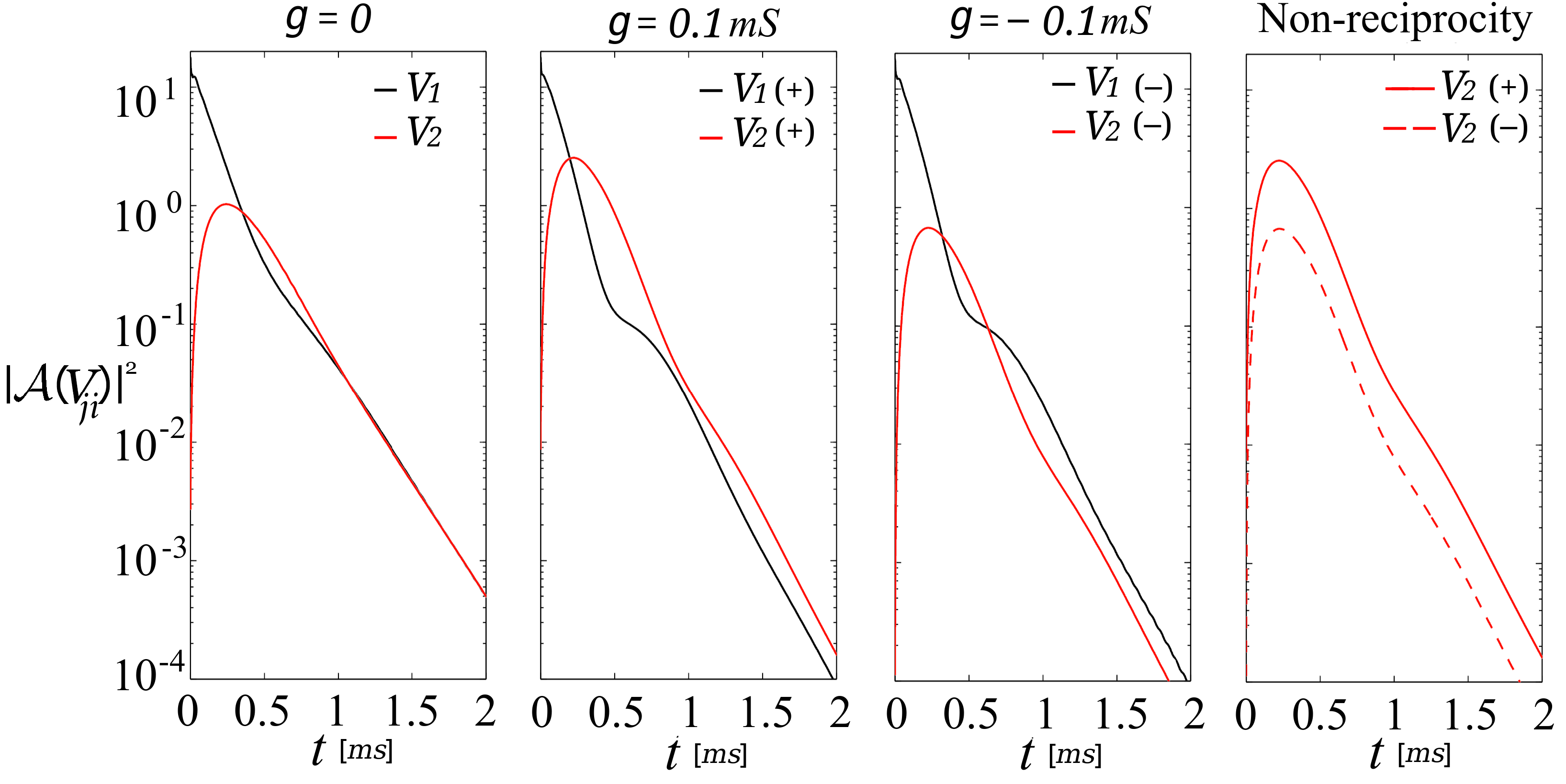}}
\caption{\footnotesize{Simulated envelope squared port voltages for different gyrator values.
\label{Impl2}}}
\end{figure}

\begin{figure}[!h]
\centerline{\includegraphics[totalheight=5.5cm]{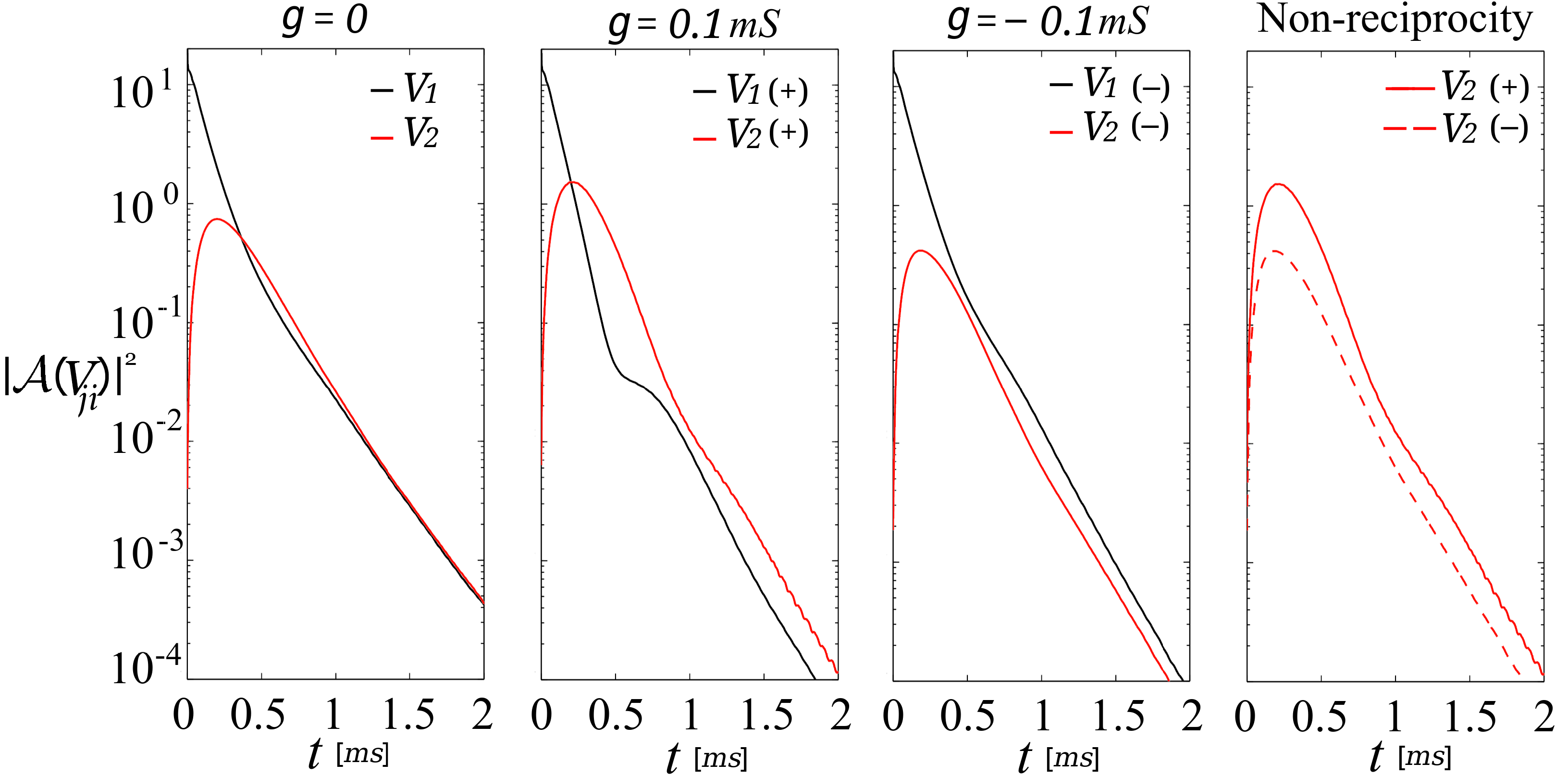}}
\caption{\footnotesize{Experimental values of envelope squared  port voltages under different gyrator conditions.\label{Impl3}}}
\end{figure}

The similar shape of the three first graphs in Fig. \ref{Impl2} and \ref{Impl3} (labeled $g=0$; $g=0.1$mS and $g=-0.1$mS) shows the disappearance of the initial neutron kaon ($K^0$ or $\bar{K}^0$) with time and the appearance of the corresponding antiparticle. Consequently, the fourth graph in each figure subsumes the $CP$ violating effect measured by the difference in the probabilities when one starts with $K^0$ or with $\bar{K}^0$. After the transient this difference is practically constant in time.

The effect due to parasitic resistances is evident in the attenuation slope. Good agreement between simulated and experimental data is evident, showing the feasibility of the proposed design approach. It should be noted that, with the actual circuit parameters, values of $\frac{\omega_a}{\omega_0}=0.2$ and $\xi=0.2$  are obtained. Although these values are not so consistent with the real kaon system as to perform precise measurements, the general behavior of both cases is similar and therefore can be easily observed.

We would like to stress that the Figs. \ref{Impl2} and \ref{Impl3} illustrate the physical realization in terms of electric networks of the mathematical equivalence between the $\mathrm{Schr\ddot{o}dinger}$ dynamics of a quantum system with a finite number of basis states and a classical dynamics.

\section{Conclusions}

The previously obtained equivalence, stricto sensu, between the $\mathrm{Schr\ddot{o}dinger}$ dynamics of a quantum system with a finite number of basis states and the classical dynamics of electric networks; namely, the isomorphism that connects in a univocal way both dynamical systems, was numerically simulated and physically realized in terms of electric circuits. This realization of the equivalence between the neutral kaon system and a classical dynamics was concreted in terms of simple circuits including gyrators in the case of $CP$ violation while maintaining the validity of $CPT$ symmetry. The comparison between dynamics implied a decomplexification procedure. The observable related to the violation of $T$ invariance at the quantum level is associated, in our realization, to the conductance of a gyrator, the two-port, non-reciprocal, passive network without loses that violates the classical symmetry $T$. The network, completely equivalent to the kaon system, allows one to represent the relevant parameters of the quantum system in terms of circuit components. In a sense, the gyrator is an equivalent representation of the weak interaction Hamiltonian.

The concept of circuit duality \cite{Bala} allows to obtain two equivalent electrical representations of the same classical differential equation, used in \cite{Caruso:2011tu}. This facilitates the selection of the parameters that govern the $CP$ or $T$ violation in the network. Moreover, there exists a one to one relationship between the states $|K^0\rangle$ and $|\bar{K}^0\rangle$ and port voltages, or currents, of the electric network. The interaction between both $LC$ subnetworks gives rise to a shift in the proper initial free frequencies, in the same way as the masses of kaons do. Moreover, the presence of proper relaxation times of the circuit are associated the mean lives $K-$\textit{short} and $K-$\textit{long}.

Analogies have always been important tools for gaining insight into physical problems, potentiated when these analogies have the character of equivalence. By analyzing the equivalent electric circuit one can improve the understanding of the $CP$ violation mechanism in kaons. For example, inspired by the present results, the connection between the the Jarlskog invariant of the three generations Cabibbo–Kobayashi–Maskawa matrix and the Berry geometrical phase is being analyzed. Other aspect of the kaon physics that could eventually be studied in terms of electric circuits is related to the different decay channels. Of course, an interested reader could go ahead with other physical ideas in both directions.

\section{Acknowledgments}

We warmly thank Prof. Jonathan Rosner for his important comments.

\end{document}